\documentclass[conference]{IEEEtran}
\usepackage{amsmath}
\usepackage{latexsym}
\usepackage{graphicx}
\usepackage{bbding}
\usepackage{indentfirst}
\usepackage{cases}
\usepackage{algorithm2e}
\usepackage{subeqnarray}
\usepackage{color}
\IEEEoverridecommandlockouts

\begin{document}
\title{Energy-Efficient Optimization for Physical Layer Security in Multi-Antenna Downlink Networks with QoS Guarantee}
\author{\authorblockN{Xiaoming~Chen and Lei~Lei
\thanks{This work was supported by the grants from the NUAA Research
Funding (No. NN2012004), the open research fund of National Mobile
Communications Research Laboratory£¬Southeast University (No.
2012D16), the Natural Science Foundation Program of China (No.
61100195) and the Doctoral Fund of Ministry of Education of China
(No. 20123218120022).}
\thanks{Xiaoming~Chen (e-mail: {\tt chenxiaoming@nuaa.edu.cn}) and
Lei~Lei are with the College of Electronic and Information
Engineering, Nanjing University of Aeronautics and Astronautics, and
also with the National Mobile Communications Research Laboratory,
Southeast University, China.}}} \maketitle

\begin{abstract}
In this letter, we consider a multi-antenna downlink network where a
secure user (SU) coexists with a passive eavesdropper. There are two
design requirements for such a network. First, the information
should be transferred in a secret and efficient manner. Second, the
quality of service (QoS), i.e. delay sensitivity, should be take
into consideration to satisfy the demands of real-time wireless
services. In order to fulfill the two requirements, we combine the
physical layer security technique based on switched beam beamforming
with an energy-efficient power allocation. The problem is formulated
as the maximization of the secrecy energy efficiency subject to
delay and power constraints. By solving the optimization problem, we
derive an energy-efficient power allocation scheme. Numerical
results validate the effectiveness of the proposed scheme.
\end{abstract}


\section{Introduction}
Without doubt, information security is a critical issue of wireless
communications due to the open nature of wireless channel.
Traditionally, information security is realized by using
cryptography technology. In fact, information theory has proven that
if the eavesdropper channel is degraded, secure communication can be
guaranteed by only using physical layer technology, namely physical
layer security \cite{Wyner} \cite{PLS1}, even if the eavesdropper
has strong computation capabilities.

The essence of physical layer security is to maximize the secrecy
rate. If there are multiple antennas at the information source,
transmit beamforming can be developed to improve the legitimate
channel capacity and to degrade the eavesdropper channel capacity,
so the achievable secrecy rate is increased \cite{Multiantenna1}
\cite{Multiantenna2}. In \cite{Multiantenna4}, the problem of
optimal beamforming was addressed by maximizing the secrecy rate. A
potential drawback of the above approach lies in that the source
must have full channel state information (CSI) to design the
transmit beam. To alleviate the assumption, a joint power allocation
and beamforming scheme was proposed based on the full CSI of the
legitimate channel and the partial CSI of the eavesdropper channel
\cite{Multiantenna5}. Yet, the CSI is difficult to obtain for the
source, especially the CSI of the eavesdropper channel. It is proved
that if there is no CSI of the eavesdropper channel, the beamforming
along the direction of the legitimate channel is optimal
\cite{CSIquantization}. Thus, the authors in \cite{CSIquantization}
proposed to convey the quantized CSI of the legitimate channel for
beamforming, and derive a performance upper bound as a function of
the feedback amount. Since the source has no knowledge of the
eavesdropper channel that varies randomly, it is impossible to
provide a steady secrecy rate. In this case, the secrecy outage
capacity is adopted as a useful and intuitive metric to evaluate
security, which is defined as the maximum rate under the constraint
that the outage probability that the real transmission rate is
greater than secrecy capacity is equal to a given value
\cite{Outagecapacity}. In multi-antenna systems, the switched beam
beamforming is a popular limited feedback beamforming scheme because
of its low complexity, small overhead and good performance
\cite{QoS1}. In this study, we propose to adopt the switched beam
beamforming to increase the capacity of the legitimate channel, and
success in deriving the closed-formed expression of the outage
probability in terms of the secrecy outage capacity.

Recently, the increasing interest in various advanced wireless
services results in an urgent demand for the communications with
quality of service (QoS) guarantee, such as delay sensitivity for
the video transmission \cite{QoS2}. Many previous analogous works
focus on the maximization of secrecy outage capacity with the QoS
requirement, in which it is optimal to use the maximum available
power. Lately, energy-efficient wireless communication
\cite{Green1}, namely green communications receives considerably
attentions due to energy shortage and greenhouse effect. In
\cite{EnergyEfficientPLS}, energy-efficient resource allocation for
secure OFDMA downlink network was studied. Therefore, we also expect
to maximize the secrecy energy efficiency, namely the number of
transmission bits per Joule, while meeting delay and power
constraints. By solving this problem, we derive an energy-efficient
adaptive power allocation scheme according to the channel condition
and performance requirement.

The rest of this letter is organized as follows. We first give an
overview of the secure multi-antenna network in Section II, and then
derive an energy-efficient power allocation scheme by maximizing the
secrecy energy efficiency while satisfying the delay and power
constraints in Section III. In Section IV, we present some numerical
results to validate the effectiveness of the proposed scheme.
Finally, we conclude the whole paper in Section V.

\section{System Model}
\begin{figure}[h] \centering
\includegraphics [width=0.45\textwidth] {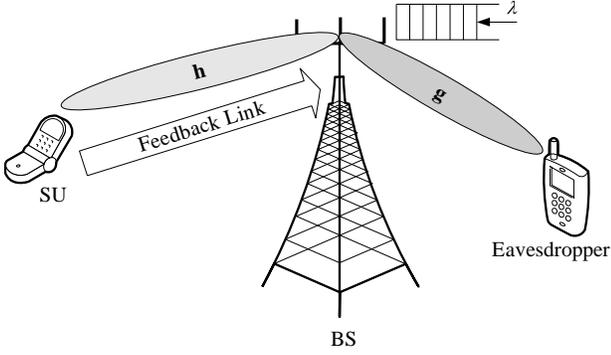}
\caption {An overview of the considered system model.} \label{Fig1}
\end{figure}

We consider a multi-antenna downlink network, where a base station
(BS) with $N_t$ antennas communicates with a single antenna secure
user (SU), while a single antenna eavesdropper also receives the
signal from the BS and tries to detect it. We use
$\alpha_s\textbf{h}$ to denote the $N_t$ dimensional legitimate
channel vector from the BS to the SU, where $\alpha_s$ is the
channel large-scale fading component, including path loss and shadow
fading, and $\textbf{h}$ represents the channel small-scale fading
component, which is a circularly symmetric complex Gaussian (CSCG)
random vector with zero mean and unit variance. Similarly, we use
$\alpha_e\textbf{g}$ to denote the $N_t$ dimensional eavesdropper
channel vector from the BS to the eavesdropper, where $\alpha_e$ and
$\textbf{g}$ are the large-scale and small-scale channel fading
components, respectively. The network is operated in time slots. We
assume that $\alpha_s$ and $\alpha_e$ remain constant during a
relatively long time period due to their slow fading, while
$\textbf{h}$ and $\textbf{g}$ remain constant in a time slot and
independently fade slot by slot. At the beginning of each time slot,
the SU selects an optimal column vector from a predetermined
$N_t\times N_t$ unitary matrix
$\mathcal{W}=\{\textbf{w}_1,\textbf{w}_2,\cdots,\textbf{w}_{N_t}\}$,
where $\textbf{w}_i$ is the $i$-th column vector, according to the
following criteria:
\begin{equation}
i^{\star}=\arg\max\limits_{1\leq i\leq
N_t}|\textbf{h}^H\textbf{w}_i|^2\label{eqn1}
\end{equation}
Then, the SU conveys the index $i^{\star}$ to the BS via the
feedback link, and the BS performs beamforming to the predetermined
signal by using $\textbf{w}_{i^{\star}}$, namely switched beam
beamforming. Thus, the receive signals at the SU and the
eavesdropper are given by
\begin{equation}
y_s=\sqrt{P}\alpha_s\textbf{h}^H\textbf{w}_{i^{\star}}x+n_s\label{eqn2}
\end{equation}
and
\begin{equation}
y_e=\sqrt{P}\alpha_e\textbf{g}^H\textbf{w}_{i^{\star}}x+n_e\label{eqn3}
\end{equation}
respectively, where $x$ is the Gaussian distributed transmit signal
with unit variance, $P$ is the transmit power, $n_s$ and $n_e$ are
the additive Gaussian white noises with unit variance at the SU and
the eavesdropper, respectively. Hence, the capacities of the
legitimate and eavesdropper channels can be expressed as
\begin{equation}
C_s=W\log_2(1+\gamma_s)\label{eqn4}
\end{equation}
and
\begin{equation}
C_e=W\log_2(1+\gamma_e)\label{eqn5}
\end{equation}
where $W$ is the spectrum bandwidth,
$\gamma_s=P\alpha_s^2|\textbf{h}^H\textbf{w}_{i^{\star}}|^2$ and
$\gamma_e=P\alpha_e^2|\textbf{g}^H\textbf{w}_{i^{\star}}|^2$ are the
signal-to-noise ratio (SNR) at the SU and the eavesdropper,
respectively. Therefore, from the perspective of information theory,
the secrecy capacity is given by $C_{sec}=[C_s-C_e]^{+}$, where
$[x]^{+}=\max(x,0)$. Since there are no knowledge of the
eavesdropper channel at the BS, it is impossible to provide a steady
secrecy capacity. In this letter, we take the secrecy outage
capacity $R_{sec}$ as the performance metric, which is defined as
the maximum rate under the condition that the outage probability
that the transmission rate surpasses the secrecy capacity is equal
to a given value $\varepsilon$, namely
\begin{equation}
P_r\bigg(R_{sec}>C_s-C_e\bigg)=\varepsilon\label{eqn7}
\end{equation}
Substituting (\ref{eqn4}) and (\ref{eqn5}) into (\ref{eqn7}), the
outage probability can be transformed as
\begin{eqnarray}
\varepsilon&=&P_r\bigg(\gamma_s<2^{R_{sec}/W}(1+\gamma_e)-1\bigg)\nonumber\\
&=&\int_0^{\infty}\int_{0}^{2^{R_{sec}/W}(1+\gamma_e)-1}f_{\gamma_{s}}(x)f_{\gamma_{e}}(y)dxdy\nonumber\\
&=&\int_0^{\infty}F_{\gamma_{s}}\left(2^{R_{sec}/W}(1+y)-1\right)f_{\gamma_{e}}(y)dy\label{eqn8}
\end{eqnarray}
where $f_{\gamma_{e}}(y)$ is the probability density function (pdf)
of $\gamma_e$, $f_{\gamma_{s}}(x)$ and $F_{\gamma_{s}}(x)$ are the
pdf and cumulative distribution function (cdf) of $\gamma_s$,
respectively. Since $\textbf{w}_{i^{\star}}$ is independent of
$\textbf{g}$, $|\textbf{g}^H\textbf{w}_{i^{\star}}|^2$ is
exponentially distributed. Thus, we have
\begin{equation}
f_{\gamma_{e}}(y)=\frac{1}{P\alpha_e^2}\exp\left(-\frac{y}{P\alpha_e^2}\right)\label{eqn9}
\end{equation}
Similarly, $|\textbf{h}^H\textbf{w}_{i^{\star}}|^2$ can be
considered as the maximum one of $N_t$ independent exponentially
distributed random variables caused by beam selection, so we have
\begin{equation}
F_{\gamma_{s}}(x)=\left(1-\exp\left(-\frac{x}{P\alpha_s^2}\right)\right)^{N_t}\label{eqn10}
\end{equation}
Substituting (\ref{eqn9}) and (\ref{eqn10}) into (\ref{eqn8}), it is
obtained that
\begin{eqnarray}
\varepsilon&=&1+\sum\limits_{n=1}^{N_t}{N_t\choose
n}(-1)^n\left(\frac{1}{1+n2^{R_{sec}/W}\alpha_e^{2}/\alpha_s^{2}}\right)\nonumber\\
&*&\exp\left(-\frac{n(2^{R_{sec}/W}-1)}{P\alpha_s^2}\right)\nonumber\\
&=&G(R_{sec},P)\label{eqn11}
\end{eqnarray}
Intuitively, $G(R_{sec},P)$ is a monotonically increasing function
of $R_{sec}$ and a monotonically decreasing function of $P$. Thus,
given transmit power $P$ and the requirement of outage probability
$\varepsilon$, the secrecy outage capacity can be derived by
computing $R_{sec}=G^{-1}(\varepsilon,P)$, where
$G^{-1}(\varepsilon,P)$ is the inverse function of $G(R_{sec},P)$.
Moreover, from (\ref{eqn11}), we can obtain the probability of
positive secrecy probability as
\begin{eqnarray}
P_r(C_{sec}>0)&=&1-G(0,P)\nonumber\\
&=&1-\frac{\alpha_s^2}{\alpha_e^2}B\left(\frac{\alpha_s^2}{\alpha_e^2},N_t+1\right)\label{eqn12}
\end{eqnarray}
where $B(x,y)$ is the Beta function. Interestingly, it is found that
the probability $P_r(C_{sec}>0)$ is independent of $P$. As $N_t$
increases, the probability increases accordingly, because more array
gains can be obtained from the switched beam beamforming.
Furthermore, (\ref{eqn12}) reveals that the short accessing distance
of the SU is benefit to enhance the information secrecy.

Since most of wireless services are delay sensitive, we take the
delay constraint into consideration. It is assumed that the data for
the SU is arrived in the form of packet of fixed length $N_b$ bits
with average arrival rate $\lambda$ (packets per slot), and it has
the minimum average delay requirements $D$ (slots) related to its
service style. Following \cite{Delayperformance}, in order to
satisfy the delay constraint, the secrecy outage capacity must meet
the following condition:
\begin{equation}
R_{sec}\geq\frac{2D\lambda+2+\sqrt{(2D\lambda+2)^2-8D\lambda}}{4D}\frac{N_b}{T}=C_{\min}\label{eqn13}
\end{equation}
where $T$ is the length of a time slot.

\section{Energy-Efficient Power Allocation}
Considering the limitation of energy resource and the requirement of
green communication, energy efficiency becomes an important
performance metric in wireless communications. In this section, we
attempt to derive a power allocation scheme to maximize the secrecy
energy efficiency while satisfying the delay and power constraints,
which is equivalent to the following optimization problem:
\begin{eqnarray}
J_1: \max&& \frac{R_{sec}}{P_0+P}\label{eqn14}\\
\textmd{s.t.}&& G(R_{sec},P)\leq\varepsilon\label{eqn15}\\
&&R_{sec}\geq C_{\min}\label{eqn16}\\
&&P\leq P_{\max}\label{eqn17}
\end{eqnarray}
where $P$ is the power consumption in the power amplifier, and $P_0$
is the power for the work regardless of information transmission,
such as the circuit power. (\ref{eqn14}) is the so called energy
efficiency, defined as the number of transmission bits per Joule.
(\ref{eqn15}) is used to fulfill the secrecy requirement based on
physical layer security, and (\ref{eqn16}) is the delay constraint,
where $C_{\min}$ is determined by both the data arrival rate
$\lambda$ and the delay requirement $D$ as shown in (\ref{eqn13}).
$P_{\max}$ is the constraint on maximum transmit power. Since
$G(R_{sec},P)$ is a monotonically increasing function of $R_{sec}$
and a decreasing function of $P$, the condition of
$G(R_{sec},P)=\varepsilon$ is optimal in the sense of maximizing the
secrecy energy efficiency. Thus, (\ref{eqn15}) can be canceled and
$R_{sec}$ can be replaced by $G^{-1}(\varepsilon,P)$. Notice that
there may be no feasible solution for $J_1$, due to the stringent
secrecy and delay constraints. Under such a condition, in order to
obtain a solution, we have to relax the constraint on the outage
probability, average delay or transmit power.

The objective function (\ref{eqn14}) in a fractional program is a
ratio of two functions of the optimization variable $P$, resulting
in $J_1$ is a fractional programming problem, which is in general
nonconvex. Following \cite{EnergyEfficientPLS}, the objective
function is equivalent to $G^{-1}(\varepsilon,P)-q^{\star}(P_0+P)$
by exploiting the properties of fractional programming, where
$q^{\star}$ is the secrecy energy efficiency when $P$ is equal to
the optimal power $P^{\star}$ of $J_1$, namely
$q^{\star}=G^{-1}(\varepsilon,P^{\star})/(P_0+P^{\star})$. Thus,
$J_1$ is transformed as
\begin{eqnarray}
J_2: \max&& G^{-1}(\varepsilon,P)-q^{\star}(P_0+P)\label{eqn18}\\
\textmd{s.t.}&& G^{-1}(\varepsilon,P)\geq C_{\min}\label{eqn19}\\
&&P\leq P_{\max}\label{eqn20}
\end{eqnarray}
$J_2$, as a convex optimization problem, can be solved by the
Lagrange multiplier method. By some arrangement, its Lagrange dual
function can be written as
\begin{eqnarray}
\mathcal{L}(\mu,\nu,P)&=&G^{-1}(\varepsilon,P)-q^{\star}(P_0+P)+\mu
G^{-1}(\varepsilon,P)\nonumber\\
&-&\mu C_{\min}-\nu P+\nu
P_{\max}\label{eqn21}
\end{eqnarray}
where $\mu\geq0$ and $\nu\geq0$ are the Lagrange multipliers
corresponding to the constraint (\ref{eqn19}) and (\ref{eqn20}),
respectively. Therefore, the dual problem of $J_2$ is given by
\begin{eqnarray}
\min\limits_{\mu,\nu}\max\limits_{P}\mathcal{L}(\mu,\nu,P)\label{eqn22}
\end{eqnarray}
For the given $\mu$ and $\nu$, the optimal power $P^{\star}$ can be
derived by solving the following KKT condition
\begin{eqnarray}
\frac{\partial\mathcal{L}(\mu,\nu,P)}{\partial
P}=(1+\mu)\frac{\partial G^{-1}(\varepsilon,P)}{\partial
P}-q^{\star}-\nu=0\label{eqn23}
\end{eqnarray}
Note that if $P^{\star}$ is negative, we should let $P^{\star}=0$.
Moreover, $\mu$ and $\nu$ can be updated by the gradient method,
which are given by
\begin{equation}
\mu(n+1)=[\mu(n)-\triangle_{\mu}(G^{-1}(\varepsilon,P)-C_{\min})]^{+}\label{eqn24}
\end{equation}
and
\begin{equation}
\nu(n+1)=[\nu(n)-\triangle_{\nu}(P_{\max}-P)]^{+}\label{eqn25}
\end{equation}
where $n$ is the iteration index, and $\triangle_{\mu}$ and
$\triangle_{\nu}$ are the positive iteration steps. Inspired by the
Dinkelbach method \cite{Dinkelbach}, we propose an iterative
algorithm as follows
\rule{8.76cm}{1pt}\\
Algorithm 1: Energy-Efficient Power Allocation\\
\rule{8.76cm}{1pt}
\begin{enumerate}

\item Initialization: Given $N_t$, $T$, $W$, $\alpha_s$, $\alpha_e$, $C_{\min}$, $P_0$,
$P_{\max}$, $\triangle_{\mu}$, $\triangle_{\nu}$ and $\varepsilon$.
Let $\mu=0$, $\nu=0$, $P=0$ and
$q^{\star}=G^{-1}(\varepsilon,P)/(P_0+P)$. $\epsilon$ is a
sufficiently small positive real number.

\item Update $\mu$ and $\nu$ according to (\ref{eqn24}) and
(\ref{eqn25}), respectively.

\item Computing the optimal $P^{\star}$ by solving the equation (\ref{eqn23}) using
some math tools, such as \emph{Mathematics} and \emph{Matlab}.

\item If
$G^{-1}(\varepsilon,P^{\star})-q^{\star}(P_0+P^{\star})>\epsilon$,
then set $q^{\star}=G^{-1}(\varepsilon,P^{\star})/(P_0+P^{\star})$,
and go to 2). Otherwise, $P^{\star}$ is the optimal transmit power.
\end{enumerate}
\rule{8.76cm}{1pt}

\section{Numerical Results}
To examine the effectiveness of the proposed energy-efficient power
allocation scheme, we present several numerical results in the
following scenarios: we set $N_t=4$, $W=10$KHz, $C_{\min}=0.8$Kb/s,
$P_0=0.5$Watt and $P_{\max}=10$Watt. $\alpha_s^2$ is normalized to1,
and we use $\rho=\alpha_e^2/\alpha_s^2$ to denote the relative
large-scale fading of the eavesdropper channel. It is found that the
proposed energy-efficient power allocation scheme converges after no
more than 20 times iterative computation in the all simulation
scenarios.

\begin{figure}[h] \centering
\includegraphics [width=0.45\textwidth] {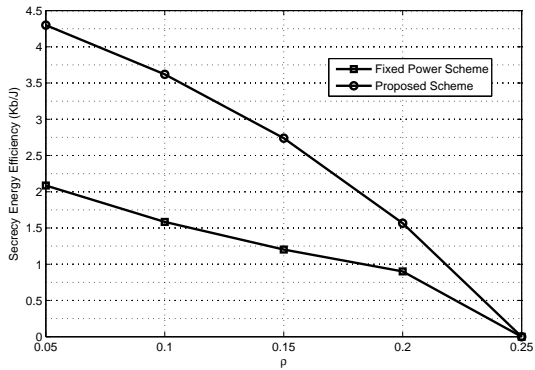}
\caption {Performance comparison of the traditional and the proposed
power allocation schemes.} \label{Fig2}
\end{figure}

Fig.\ref{Fig2} compares the secrecy energy efficiency of the
proposed adaptive power allocation scheme and the fixed power
allocation scheme with $\varepsilon=0.05$. Intuitively, it is
optimal to use $P_{\max}$ as the transmit power in the sense of
maximizing the secrecy outage capacity, so we set $P=P_{\max}$
fixedly for the fixed power allocation scheme. As seen in
Fig.\ref{Fig2}, the proposed scheme performs better than the fixed
one, especially when $\rho$ is small. For example, when $\rho=0.10$,
there is about $2$Kb/J gain. Therefore, the proposed scheme is more
suitable for the future green and secure communications. It is found
that when $\rho=0.25$, the secrecy energy efficiency reduces to
zero. This is because there is no nonzero secrecy outage capacity
under such constraint conditions. In order to support the case with
the large $\rho$, we need to relax the constraint conditions or
deploy more antennas at the BS to obtain more array gain.

\begin{figure}[h] \centering
\includegraphics [width=0.45\textwidth] {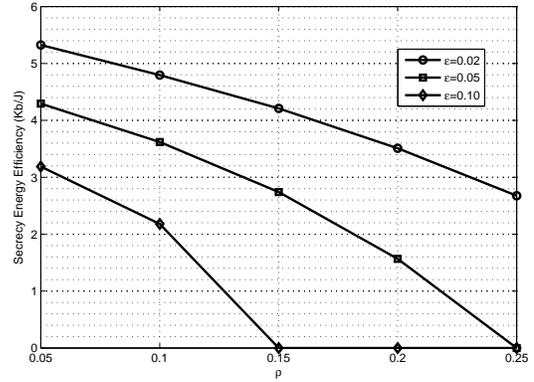}
\caption {Performance comparison of the proposed power allocation
scheme with different requirements of secrecy outage probability.}
\label{Fig3}
\end{figure}

Fig.\ref{Fig3} investigates the effect of the requirements of outage
probability on the secrecy power efficiency of the proposed scheme.
For a given $\rho$, as $\varepsilon$ decreases, the secrecy energy
efficiency reduces accordingly, this is because more power is used
to decrease the outage probability. On the other hand, for a given
requirement of the outage probability, the increase of $\rho$ leads
to the decrease of the secrecy energy efficiency, since the
eavesdropper has a strong eavesdropping ability.

\section{Conclusion}
A major contribution of this paper is the introduction of an
energy-efficient power allocation scheme into a multi-antenna
downlink network employing physical layer security with delay
guarantee. Considering the importance of the CSI in multi-antenna
networks, the switched beam beamforming is adopted to realize the
adaptive transmission. Numerical results confirm the effectiveness
of the proposed scheme. In future works, we will further study the
cases with multi-antenna eavesdropper, imperfect CSI, robust
beamforming, etc.

\end{document}